# Detection of Weak Near-Infrared Signal Based on Digital Orthogonal Vector Lock-in Amplifier

Qi-Jie Tang, Yi-hao Zhang, Shu-cheng Dong, Jin-ting Chen, Feng-xin Jiang, Zhi-yue Wang, Ya-qi Chen, Hong-fei Zhang, Jian Wang, *Senior Member, IEEE*

*Abstract*—A near-infrared (NIR) measurement based on digital orthogonal vector lock-in amplifier (LIA) is present in this paper. NIR sky background radiation is very weak. To detect the signals obscured by noise, the best way achieved is to use a chopper to modulate the detected signal and using a LIA to demodulate. The effect of 1/f noise of detector, dark current and other noises can be reduced to get sufficient signal-to-noise ratio (SNR). The orthogonal vector LIA can avoid the phase shift on the accuracy of measurement by two orthogonal components. In order to simplify the system, a digital algorithm is adopted to realize the LIA which is operated in a microcontroller with ARM cortex-M4. Data is obtained through ADC and the signal of detector is amplified and filtered. Then the phase sensitive detection (PSD), low-pass filter (LPF) and amplitude phase calculation are performed. The digital method can greatly simplify the circuit, and conveniently adjust the time constant of the LPF to realize the different equivalent noise bandwidth (ENB). The algorithm has the specification of high precision, flexible usage, simple implementation and low computation resource. By using this method, the weak infrared signal submerged by the noise can be obtained, which extremely improves the detection capability of the system.

## I. INTRODUCTION

The sky background plays an extremely important role in astronomy, which affects the entire electromagnetic spectrum observation. Infrared sky background greatly restricts the performance of infrared telescopes and other equipment, such as the depth of the sky, the limiting magnitude and the exposure time of astronomical imaging system. The average intensity and variation parameters of local infrared sky background are important references for evaluating the suitability of a candidate site. In order to make the equipment more portable, large optical aperture cannot be used which cause the actual collected energy is very limited, and the 1/f noise and dark current of detector limit the detection ability of the measurement system. Since the advent of lock-in amplifier, it shows excellent performance in weak signal detection and has been widely used in many fields of scientific research which promote the development of physical, chemical, biological medicine, earthquake, communications, astronomy, and others. For example, the molecular beam mass spectrometers (MBMS), scanning electron microscope (SEM), Auger electron spectrometer and other instruments have adopted the LIA. According to the principle and characteristics of the LIA, we design a weak infrared signal detection based on digital orthogonal vector LIA.

## II. SYSTEM DESIGN

The measurement system consists of chopper, detector, high-gain amplifier, acquisition and calculation. The structure is shown in Figure 1. The infrared radiation signal is tuned into AC signal by optical chopper to avoid the influence of 50Hz power noise and the modulation frequency is set up to about 82Hz.The optical signal is converted into an electrical signal by the infrared PIN photodiodes. After the amplification, the signal is sampled and digitized by ADC. Then the digital phase lock is implemented in the microcontroller including PSD, LPF, amplitude phase calculation and other digital LIA functions.

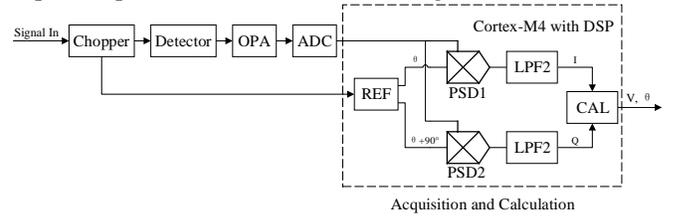

Figure 1 Structure of LIA

The output of the detector is weak current signal, which contains the DC base and AC modulation signal. The high-gain amplifier adopts the three stage structure, as shown in Figure 2.

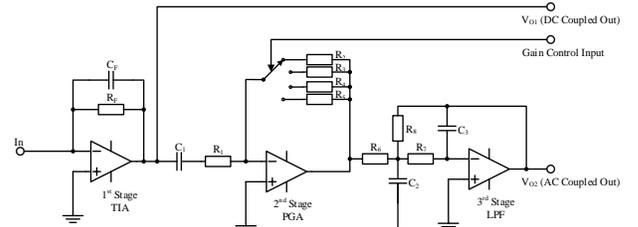

Figure 2 Structure of High-gain Amplifier

This work was supported by the National Natural Science Funds of China under Grant No: 11603023, 11773026, 11728509, the Fundamental Research Funds for the Central Universities (WK2360000003, WK2030040064), the Natural Science Funds of Anhui Province under Grant No: 1508085MA07, the Research Funds of the State Key Laboratory of Particle Detection and Electronics, the CAS Center for Excellence in Particle Physics, the Research Funds of Key Laboratory of Astronomical Optics & Technology, CAS.

The Authors Qi-Jie Tang, Yi-hao Zhang, Shu-cheng Dong, Jin-ting Chen, Feng-xin Jiang, Zhi-yue Wang, Ya-qi Chen, Hong-fei Zhang, Jian Wang is with the University of Science and Technology of China, Jian Wang, State Key Laboratory of Technologies of Particle Detection and Electronics, University of Science and Technology of China, Hefei, Anhui 230026, China (e-mail: Hong-fei Zhang, nghong@ustc.edu.cn; Jian Wang, wangjian@ustc.edu.cn).

The first stage is the DC coupling trans-resistance amplifier (TIA), which amplifies the weak current signal into voltage signal. To get the best noise performance (SNR proportional to the $\sqrt{R_F}$), the larger the feedback resistance $R_F$, the larger the gain, and the better the SNR. The selection of OPA is very important, the low-offset, low-noise JFET or CMOS architecture OPA should be done. The second stage is the AC coupling inverting input programmable-gain amplifier (PGA), which is used to eliminate DC baseline and avoid large 1/f noise of detector. The third stage is a second-order LPF, which filter the high frequency noise and limit the bandwidth according to the frequency of the chopper. The total gain can reach $10^{12} \text{V/A}_\circ$

The LIA has excellent performance in weak signal detection, and its suppression noise has three basic starting points:(1) Using the modulator to transfer the spectrum of dc or slow-changing signals to the modulation frequency ω0 to avoid the adverse effects of 1/f noise. (2) Using PSD to realize the demodulation process of modulating signal, in which the frequency ω0 and phase θ are used simultaneously, and the probability of noise and signal in the same frequency and phase is very low. (3) Using a LPF instead of a band-pass filter (BPF) to suppress broadband noise. The frequency band of LPF can be very narrow, and its bandwidth is not affected by the modulation frequency, and the stability is much better than the BPF.

There are many implementation schemes for LIA, in which the PSD is the core component and it is commonly used in analog multiplier and electronic switch. But it is more complicated to use analog circuit for its large volume and high power consumption. It is not necessary to increase the hardware resource on the digital scheme. In our design all digital processes are completed in the microcontroller with Cortex-M4 which consists of floating point unit (FPU) and rich interface and peripherals. This microcontroller can meet the requirements of real-time processing, all of the control, collection, computing, communications and other functions.

The signal-to-noise ratio of LIA is directly proportional to the time constant of the LPF. In theory, as long as the time constant of LPF is large enough, the bandwidth of the BPF can be narrow enough, thus the ability of restraining noise can be greatly improved. LIA can track the frequency of the chopper, and the fluctuation of frequency will not affect the calculation result. The digital method simplifies the system and requires only the basic amplification and data acquisition. The time constant of LPF can be adjusted conveniently to realize different equivalent noise bandwidth (ENB).

## III. TEST RESULT

The calculation results of the digital LIA of the signal source are recorded by using the sinusoidal signal source instead of the detector and the amplifier input to the ADC. In theory, the response is in line with the shape of the mean filter and is shifted to f0, the value A(f) is

$$A(f) = A_0 \left| \frac{\sin[(f-f_0)\pi T]}{(f-f_0)\pi T} \right| = A_0 |sinc[(f-f_0)\pi T]| \quad (1)$$

F0 is the modulation frequency of 82.03125 Hz; T is the time constant and set by the program; A0 is the amplitude of the signal and equal to the peak value of the signal source of 8V.The time constant T is set to 1s and 20s, the frequency response curve of the LIA is shown in Figure 3.

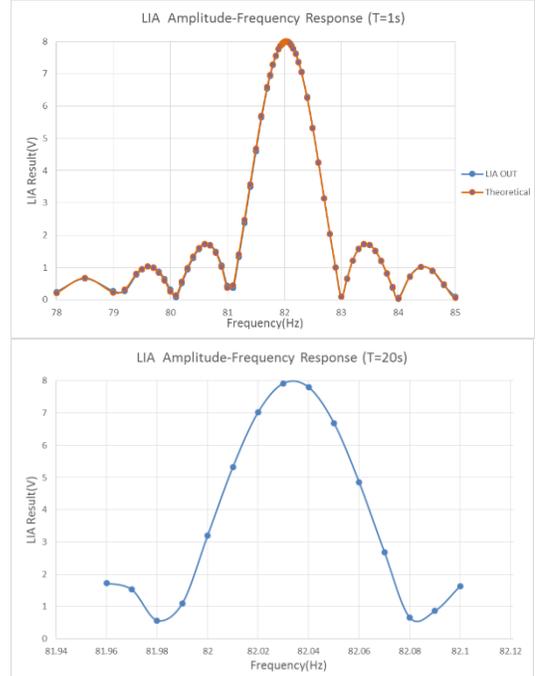

Figure 3 Frequency response of LIA

The test results are fully consistent with the theoretical values. The ENB is equal to

$$ENB = \int_{f=0}^{f=\infty} sinc^2[(f-f_0)\pi T]df = \frac{1}{T} \quad (2)$$

ENB is inversely proportional to the time constant. ENB is 1Hz for T=1s, and is 0.05 Hz when increase the time constant to 20s. The ENB of LIA is inversely proportional to the time constant, so the SNR is.

For different system requirements, flexible time constant can be used to improve the response speed of detection by using as short time constant as possible under the premise of SNR. For example, when using a low-noise InGaAs detector, the time constant can be set to 1~2s to achieve the desired SNR while using the noisy InSb detector, the time constant can be set to 5~10s.

Preliminary observations of Ngari (Ali) in Tibet were carried out using the near-infrared sky background measuring instrument based on the digital orthogonal vector lock-in amplifier. The observation data shows that the equipment can achieve sufficient resolution and sensitivity.

## IV. CONCLUSION

Weak infrared signal detection based on the digital orthogonal vector LIA has incomparable advantages compared with common methods. Furthermore, adopting digital solutions to simplify the circuit, reduce the size and power consumption of equipment and improve the integration of system and environment adaptability. The method has been used to detect the weak infrared signal with a stable and reliable operation. The SNR meets the design indexes which has great significance to the infrared astronomy observation.